\documentstyle[12pt]{article}
\def\beqa{\begin{eqnarray}}
\def\eeqa{\end{eqnarray}}
\def\beq{\begin{equation}}
\def\eeq{\end{equation}}
\begin{document}
\begin{titlepage}
   \title{Neutrino Oscillations in Caianiello's Quantum Geometry Model}
\author{V. Bozza$^{a,b}$\thanks{E-mail: valboz,capozziello,lambiase,scarpetta@sa.infn.it},
S Capozziello$^{a,b}$,
%\thanks{E-mail: capozziello@sa.infn.it},
 G. Lambiase$^{a,b}$,
 %\thanks{E-mail: lambiase@sa.infn.it},
 and G. Scarpetta$^{b,c,d}$
 %\thanks{E-mail: scarpetta@sa.infn.it}
 \\
 {\em $^a$Dipartimento  di Scienze Fisiche ``E.R. Caianiello'',} \\
 {\em  Universit\`a di Salerno, 84081 Baronissi (Sa), Italy.} \\
 {\em $^b$Istituto Nazionale di Fisica Nucleare, Sez. di Napoli, Italy.} \\
 {\em $^c$Dipartimento di Fisica, } \\
 {\em Universit\'a di Salerno, 84081 Baronissi (Sa),
 Italy}\\
 {\em $^d$IIASS, Vietri sul Mare (Sa), Italy.} }
\date{\today}
\maketitle
\begin{abstract}
Neutrino flavor oscillations are analyzed in the framework of
Quantum Geometry model proposed by Caianiello. In particular, we
analyze the consequences of the model for accelerated neutrino
particles which experience an effective Schwarzschild geometry
modified by the existence of an upper limit on the acceleration,
which implies a violation of the equivalence principle. We find a
shift of quantum mechanical phase of neutrino oscillations, which
depends on the energy of neutrinos as $E^3$. Implications on
atmospheric and solar neutrinos are discussed.
\end{abstract}

\thispagestyle{empty}

\vspace{10.mm}


\vfill

\end{titlepage}

\section{Introduction}

The long--standing problem about the deficiency of the solar
neutrino and the atmospheric neutrino problem might be explained
invoking oscillations between the various flavors or generations
of neutrinos. In fact, neutrino oscillations can occur in the
vacuum if the eigenvalues of the mass matrix are not all
degenerate, and the corresponding mass eigenstates are different
from weak interaction eigenstates (Bilenky and Pontecorvo, 1978).
The most discussed version of this type of solutions is the MSW
effect (Mikhyev and Smirnov, 1986a, 1986b; Wolfeinstein, 1978) in
which solar electron neutrinos are converted almost completely in
muon or tau neutrinos owing to the presence of matter in the Sun.

Recently, the quantum mechanical oscillations of neutrinos
propagating in a gravitational field (usually the Schwarzschild
field) have been discussed by several authors (see for example
(Ahluwalia and Burgad, 1996; Bhattacharya {\it et al.}, 1999) and
reference therein), also in view of astrophysical consequences.
Ahluwalia and Burgard consider in fact, the gravitational effect
on the oscillations showing that an external weak gravitational
field of a star adds a new contribution to the phase difference
(Ahluwalia and Burgard, 1996). They also suggest that the new
oscillation phase may be a significant effect on the supernova
explosions since the extremely large fluxes of neutrinos are
produced with different energies corresponding to the flavor
states. This result has been also discussed by Bhattachya, Habib
and Mottola (Bhattacharya {\it et al.}, 1999). In their approach
it is shown that the possible gravitational effect appears at the
higher order with respect to one calculated in Ref. (Ahluwalia and
Burgard, 1996), with a magnitude of the order $10^{-9}$, which is
negligible in typical astrophysical applications.

An alternative mechanism of neutrino oscillations has been
proposed in (Gasperini, 1998, 1989; Halprin and Leung, 1991) as a
means to test the equivalence principle. In this mechanism,
neutrino oscillations follow by assuming a flavor non--diagonal
coupling of neutrinos to gravity which violates the equivalence
principle, i.e. if the universality of the gravitational
couplings to different flavors breaks down, additional phase
difference appears. Therefore, the understanding how the presence
of a gravitational field or the violation of the equivalence
principle affects the neutrino oscillation phase is an important
matter.

In this paper we face this issue in the framework of Quantum
Geometry model proposed by Caiainiello some years ago in the
attempt to unify Quantum Mechanics and General Relativity
principles (see (Caianiello, 1981, 1992) and references therein).
In this model the effective four--dimensional metric depends on
the mass of a given test particle, so that {\it test particles
with different rest masses experience different geometries and, as
consequence, an effective violation of the equivalence principle
occurs}. The geodesic paths along which the test particles are
moving become mass--dependent, resulting in a non--universality of
the gravitational coupling (Caianiello {\it et al}, 1990),  and
making the metric observer--dependent, as also conjectured by
Gibbons and Hawking (Gibbons and Hawking, 1977).

The view frequently held that the proper acceleration of a
particle is limited upwardly (Caianiello {\it et al.}, 1982) finds
in this model a geometrical interpretation epitomized by the line
element
\begin{equation}\label{1}
d\tilde{s}^2= \left(1+\frac{g_{\mu\nu}d\ddot{x}^\mu
d\ddot{x}^\nu}{{\cal A}_m^2}\right)ds^2\equiv \sigma^2 (x)
ds^2\,{,}
\end{equation}
experienced by the accelerating particle along its worldline. In
(\ref{1}) ${\cal A}_m=2mc^3/\hbar$ is the proper Maximal
Acceleration (MA) of the particle of mass $m$ and $\ddot{x}^\mu$
its four--acceleration.

MA has several implications. It provides a regularization method
in Quantum Field Theory (Feoli {\it et al.}, 1999b), allowing to
circumvent inconsistencies associated with the application of the
point-like concept to relativistic quantum particles, it is the
same cut--off on the acceleration required in an {\it ad hoc}
fashion by Sanchez in order to regularize the entropy and the free
energy of quantum strings (Sanchez, 1993), and it is also invoked
as a necessary cut--off by McGuigan in the calculation of black
hole entropy (McGuigan 1994).

Applications of Caianiello's model include cosmology (Caianiello
{\it et al.}, 1991; Capozziello {\it et al.}, 1999), where the
initial singularity can be avoided while preserving inflation, the
dynamics of accelerated strings (Feoli, 1993) and the energy
spectrum of a uniformly accelerated particle (Caianiello {\it et
al.}, 1990).

The extremely large value that ${\cal A}_m$ takes for all known
particles makes a direct test of the model difficult. Nonetheless
a direct test that uses photons in a cavity has also been
suggested (Papini {\it et al.}, 1995). More recently, we have
worked out the consequences of the model for the classical
electrodynamics of a particle (Feoli {\it et al.}, 1997), the mass
of the Higgs boson (Lambiase {\it et al.}, 1999; Kuwata, 1996) and
the Lamb shift in hydrogenic atoms (Lambiase {\it et al.}, 1998).
In the last instance, the agreement between experimental data and
MA corrections is very good for $H$ and $D$. For $He^+$ the
agreement between theory and experiment is improved by $50\%$
when MA corrections are included. MA effects in muonic atoms
appear to be measurable in planned experiments (Chen {\it et
al.}, 1999). MA also affects the helicity and chirality of
particles (Chen {\it et al.}, 2000). Very recently the behaviour
of classical (Feoli, {\it et al.}, 1999a) and quantum (Capozziello
{\it et al.}, 2000a) particles in a Schwarzschild field with MA
modifications have been studied.

A limit on the acceleration also occurs in string theory. Here the
upper limit manifests itself through Jeans-like instabilities
(Sanchez and Veneziano, 1990; Gasperini {\it et al.}, 1991)  which
occur when the acceleration induced by the background
gravitational field is larger than a critical value $a_c =
(m\alpha)^{-1}$for which the string extremities become causally
disconnected (Gasperini, 1992). $m$ is the string mass and
$\alpha$ is the string tension. Frolov and Sanchez (Frolov and
Sanchez, 1991) have then found that a universal critical
acceleration $a_c$ must be a general property of strings. It is
worth to note that it is possible to derive, in the framework of
Caianiello's Quantum Geometry model, the generalized uncertainty
principle of string theory (Capozziello {\it et al.}, 2000b).

The paper is organized as follows. In Section 2 we shortly discuss
the Quantum Geometry model and derive the modified Schwarzschild
geometry by taking into account the MA corrections (for details
see (Feoli {\it et al.}, 1999a)). In Section 3 we calculate the
corrections induced by MA to the quantum mechanical phase of mixed
states of neutrinos radially propagating in the modified
Schwarzschild geometry. Conclusions are drawn in Sections 4.

\section{Modified Schwarzschild Space-Time \\ in Quantum Geometry}

The model proposed by Caianiello, which includes the effects of MA
in dynamics of particles, was to enlarge the space-time manifold
to an eight-dimensional space-time tangent bundle $TM_8$. In this
way the invariant line element is defined as (Caianiello {\it et
al.}, 1990a)
\begin{equation}\label{2.1}
d\tilde{s}^{2}=g_{AB}dX^{A}dX^{B},\quad A,B=1,...,8\,{,}
\end{equation}
where the coordinates of $TM_8$ are
\begin{equation}\label{2.2}
X^{A}=\left(x^{\mu};\frac{1}{{\cal
A}_m}\frac{dx^{\mu}}{ds}\right), \quad \mu=1,...,4\,{,}
\end{equation}
and
\begin{equation}\label{2.3}
g_{AB}=g_{\mu\nu}\otimes g_{\mu\nu}\,, \quad
ds^2=g_{\mu\nu}dx^{\mu}dx^{\nu}\,{.}
\end{equation}
$ds$ is the ordinary line element of the four--dimensional
space--time and $dx^{\mu}/ds$ is the four--velocity of the
particle moving along its worldline. In Eq. (\ref{2.2}), ${\cal
A}_m$ is the MA depending, in the quantum geometry theory proposed
by Caianiello, on the mass  $m$ of the particle, whose value is
given by ${\cal A}_m=2mc^3/\hbar$. In other models, ${\cal A}_m$
is interpreted as an universal constant and $m$ is replaced by the
Plank mass $m_P$. Using Eqs. (\ref{2.2}) and (\ref{2.3}) the line
element  (\ref{2.1}) can be written as
\begin{equation}\label{2.4}
 d \tilde {s}^{2}=\left(1+\frac{g_{\mu\nu}\ddot{x}^{\mu}\ddot{x}^{\nu}}{{\cal
 A}_m^2}
 \right)g_{\alpha\beta}dx^{\alpha}dx^{\beta}\equiv
 \sigma^2(x) g_{\alpha\beta}dx^{\alpha}dx^{\beta}\,{,}
\end{equation}
where $\ddot{x}^{\mu}=d^2x^{\mu}/ds^2$ is the four--acceleration
of particles and $ds^2=g_{\mu\nu}dx^{\mu}dx^{\nu}$ is the metric
due to a background gravitational field. In the absence of
gravity, $g_{\mu\nu}$ is replaced by the Minkowski metric tensor
$\eta_{\mu\nu}$. The embedding procedure has been developed to
find the effective space-time geometry in which a particle can
move when the constraint of a MA is present (Caianiello {\it al.},
1990b). In fact, if one finds the parametric equations that relate
the velocity field $\dot{x}^{\mu}$ to the first four coordinates
$x^{\mu}$, one can calculate the effective four--dimensional
metric on the hypersurface locally embedded in $TM_8$. This
procedure strongly depends on the choice of the velocity field of
the particle. From Eq. (\ref{2.4}) it follows also that even
starting from a phase space $TM_8$ with a flat metric, i.e.
$g_{AB}=\eta_{\mu\nu}\otimes\eta_{\mu\nu}$, in the case of
accelerating particles characterized by a velocity field
$\dot{x}^{\mu}$ not trivially constant, one gets an effective
four--dimensional geometry which, in general, is curved. In other
words, even though the background space--time is flat, the
effective geometry experienced by an accelerating particle is
curved.

We stress that the curvature of the effective geometry is not
induced by matter through the conventional Einstein equations: It
is due to the motion in the momentum space and vanishes in the
limit $\hbar \to 0$. Thus, it represents a quantum correction to
the given background geometry.

In order to calculate the corrections to the Schwarzschild field
experienced by a particle initially at infinity and falling toward
the origin along a geodesic, one must calculate the metric induced
by the embedding procedure (\ref{2.4}). On choosing $\theta
=\pi/2$, one finds the conformal factor produced by the embedding
procedure
 \begin{equation}\label{eq4}
\sigma^2(r)=1+\frac{1}{{\cal A}^2_m}
\left[\left(1-\frac{2M}{r}\right)\ddot{t}^{\, 2}-
\frac{\ddot{r}^2}{1-2M/r}-r^2\ddot{\phi}^2\right]\,{,}
\end{equation}
where $\ddot{t},\ddot{r}$ and $\ddot{\phi}$ are given by the
standard results (Misner {\it et al.}, 1973)
 \begin{eqnarray}
\ddot{t}^2 & = &\frac{\tilde{E}^2}{(1-2M/r)^4}\frac{4M^2}{r^4}
\left[\tilde{E}^2-\left(1-\frac{2M}{r}\right)\left(1+\frac{\tilde{L}^2}{r^2}
 \right)\right]\,{,} \nonumber \\
\ddot{r}^2 & = &\left(-\frac{M}{r^2}+\frac{\tilde{L}^2}{r^3}-
\frac{3M\tilde{L}^2}{r^4}\right)^2\,{,}\label{eq5} \\
 \ddot{\phi}^2 & = &
\frac{4\tilde{L}^2}{r^6}\left[\tilde{E}^2-
\left(1-\frac{2M}{r}\right)\left(1+\frac{\tilde{L}^2}{r^2}
\right)\right]\,{.} \nonumber
\end{eqnarray}
$M$ is the mass of the source, $\tilde{E}$ and $\tilde{L}$ are the
total energy ($E$) and angular momentum ($L$) per unit of particle
mass $m$. The conformal factor $\sigma^2(r)$ is then given by
(Feoli {\it et al.}, 1999a)
 $$
\sigma^2(r)=1+\frac{1}{{\cal A}_m^2}\left\{
-\frac{1}{1-2M/r}\left(-\frac{3M\tilde{L}^2}{r^4}+\frac{\tilde{L}^2}{r^3}
-\frac{M}{r^2}\right)^2 + \right.
 $$
 \begin{equation}\label{eq6}
 \left.
+\left(-\frac{4\tilde{L}^2}{r^4}+\frac{4\tilde{E}^2
M^2}{r^4(1-2M/r)^3}
\right)\left[\tilde{E}^2-\left(1-\frac{2M}{r}\right)\left(1+
\frac{\tilde{L}^2}{r^2}\right)\right]\right\}\,{.}
\end{equation}
Modifications to the Schwarzschild geometry experienced by
radially ($\tilde{L}=0$) accelerating neutrinos are easily
calculated. In fact, from Eq. (\ref{eq6}) and by using the weak
field approximation, one gets
\begin{equation}\label{2}
\sigma^2(r)=1-\frac{1}{{\cal
A}_m^2}\left(\frac{1}{4}+\frac{E^2}{m^2}
-\frac{E^4}{m^4}\right)\frac{r_s^2}{r^4}\,,
\end{equation}
where $r_s=2GM/c^2$ is Schwarzschild radius.

\section{MA Corrections to Quantum Mechanical Phase}

Corrections induced by MA to the quantum mechanical phase mixing
of massive neutrinos are calculated following Ref. (Bhattacharya
{\it et al.}, 1999). In the semiclassical approximation, i.e. the
action of a particle is considered as a quantum phase, a particle
propagating in a gravitational field from a point A to a point B,
changes its quantum mechanical phase according to the relation
(Stodolski, 1979)
\begin{equation}\label{qmp}
\Phi =\frac{1}{\hbar}\int_A^B md\tilde{s}=\frac{1}{\hbar}\int_A^B
p_{\mu}dx^{\mu}\,.
\end{equation}
Here $p_{\mu}=m\tilde{g}_{\mu\nu}(dx^{\nu}/d\tilde{s})$ is the
four--momentum of the particle and $\tilde{g}_{\mu\nu}=\sigma^2(r)
g_{\mu\nu}$, where the conformal factor $\sigma^2(x)$ is defined
in Eq. (\ref{2}). In order that different neutrinos could
interfere at the same final point B, with coordinates $(t_B,
r_B)$, one requires, in the geometrical optical approximation,
that the relevant components of the wave function have not started
from the same initial point A, with coordinates $(t_A, r_A)$.
Then, the quantum mechanical phase becomes
\begin{equation}\label{qmpr}
\Phi =\frac{1}{\hbar}\int_{r_A}^{r_B} p_{r}dr\,.
\end{equation}
Inserting the momentum of the particle, calculated by mass-shell
condition $\tilde{g}^{\mu\nu}p_{\mu}p_{\nu}=m^2$,
\begin{equation}\label{moment}
 p_{r}=\frac{\sqrt{E^{2}-m^{2}\sigma^2(1-r_s/r)}}{1-r_s/r}
\end{equation}
into Eq. (\ref{qmpr}) one gets, up to second order in $r_s/r$,
 \begin{equation}\label{qmpf}
 \Phi =\Phi_0+\Phi_{{\cal A}_m}
 \end{equation}
 where
 \begin{equation}\label{mot}
 \Phi_0=\frac{\sqrt{E^2-m^2}}{\hbar}(r_B-r_A)+\frac{(2E^2-m^2)r_s}{2\sqrt{E^2-m^2}}
 \log\frac{r_s}{r}+
 \end{equation}
 $$
 -\frac{r_s^2\sqrt{E^2-m^2}}{\hbar}\left(1+\frac{m^2}{2(E^2-m^2)}
 +\frac{m^4}{8(E^2-m^2)^2}\right)\left(\frac{1}{r_B}-\frac{1}{r_A}\right)
 \,{,}
 $$
 represents the result of Ref. (Bhattacharya {\it et al.}, 1999), and
\begin{equation}\label{qmpMA}
 \Phi_{{\cal A}_m}=\frac{1}{{\cal A}_m^2}\left(\frac{1}{4}+
 \frac{E^2}{m^2}-\frac{E^4}{m^4}\right)\frac{m^2r_s^2}{6\hbar\sqrt{E^2-m^2}}
 \left(\frac{1}{r_B^3}-\frac{1}{r_A^3}\right)
\end{equation}
is the contribution due to the MA. For ultra--relativistic
neutrinos, $E>>m$, the relative quantum mechanical phase
$\Delta\Phi$ of the two different mass eigenstates is given by
\begin{equation}\label{phase}
 \Delta\Phi=\Delta\Phi_{(0)}+\Delta\Phi_{{\cal A}_m}\,,
\end{equation}
where
\begin{equation}\label{phasemot}
 \Delta\Phi_{(0)}=\frac{\Delta m^2}{2E\hbar}(r_B-r_A)+
 \frac{\Delta m^2}{4E^2}
 (r_B-r_A)-\frac{\Delta m^2(m_1^2+m_2^2)r_s}{8\hbar
 E^3}\log\frac{r_B}{r_A}\,{,}
\end{equation}
as in Ref. (Bhattacharya {\it et al.}, 1999), and
\begin{equation}\label{phaseMA}
 \Delta\Phi_{{\cal A}_m}=\frac{\hbar E^3}{24}
 \frac{\Delta m^2(m_1^2+m_2^2)}{m_1^4m_2^4}
 \frac{r_s^2(r_B^3-r_A^3)}{r_B^3r_A^3}\,{.}
\end{equation}
Here $\Delta m^2=\vert m_2^2-m_1^2\vert$. In Eq. (\ref{phasemot}),
the first term represents the standard phase of neutrino
oscillations, the second term is the kinetic correction to the
first order, and finally, the last term is the gravitational
correction to the leading order. The second and third term in Eq.
(\ref{phasemot}) can be neglected with respect the first term, so
that we will neglect them in what follows. Notice that $\Delta
\Phi_{{\cal A}_m}\to 0$ as $\hbar\to 0$. It is more convenient to
rewrite the phases (\ref{phasemot}) and (\ref{phaseMA}) in the
following way
\begin{equation}\label{phasemot1}
\Delta\Phi_{(0)}=2.5\cdot 10^3 \frac{\Delta
m^2}{\mbox{eV}^2}\frac{\mbox{MeV}}{E}\frac{r_A-r_B}{\mbox{Km}}\,{,}
\end{equation}
and
\begin{eqnarray}
 \Delta\Phi_{{\cal A}_m}&=&2.4\cdot 10^8\frac{\Delta m^2}{\mbox{eV}^2}
 \frac{E^3}{\mbox{MeV}^3}\frac{M^2}{M^2_{\odot}}
 \frac{\mbox{eV}^6}{(m_1m_2)^4/(m_1^2+m_2^2)} \times \nonumber \\
 && \times \frac{\mbox{Km}^3}{(r_Ar_B)^3/(r_A^3-r_B^3)} \label{phaseMA1}
 \,{,}
\end{eqnarray}
where $M_{\odot}$ is the solar mass.

Comparison between the quantum mechanical phases (\ref{phasemot1})
and (\ref{phaseMA1}) are reported in Table I for atmospheric
neutrinos with mass--squared difference $\Delta m^2=(10^{-2}\div
10^{-3})$eV$^2$. We have assumed the following numerical values:
$r_A=R_{Earth}=6.3\cdot 10^3$Km and $r_B=r_A+10$Km, $r_s\sim
10^{-6}$Km is the Schwarzschild radius for the Earth and, finally,
the energy of neutrinos is $E\sim 1$GeV. MA corrections to the
quantum mechanical phase are meaningful for neutrinos with masses
$m_1, m_2\sim 0.05\div 0.1$eV. In this range, in fact, such
corrections turn out to be $10^{-2}\div 10^{-3}$ smaller than the
phase (\ref{phasemot}).

For solar neutrinos, we have a similar situation. Results are
summarized in Table II for the values $\Delta m^2=(10^{-10}\div
10^{-12})$eV$^2$, $r_A=R_{Earth}$, $r_B=r_A+1,5\cdot 10^8$Km,
$E\sim 1$MeV and $E\sim 10$MeV, $M\sim M_{\odot}$. Again, the
quantum mechanical phase corrections induced by MA become
relevants for neutrino masses of the order $0.05\div 0.1$eV.

Masses below $0.05$eV lead to high corrections that cannot be
treated in this perturbative model.

It is worthwhile to point out the different dependence on the
energy of the two phases: $\Delta\Phi_{(0)}\sim E^{-1}$ and $
\Delta\Phi_{{\cal A}_m}\sim E^3$. This can notably help the
separation of the two components in experimental tests, because
the weight of MA corrections is largely affected by the energy of
neutrinos. A good statistical analysis could succeed in bringing
this term to light.

\section{Conclusions}

Einstein's equivalence principle plays a fundamental role in the
construction and testing of theories of gravity. Though verified
experimentally to better than a part in $10^{11}$ for bodies of
macroscopic dimensions, doubts have at times been expressed as to
its validity down to microscopic scales. It is conceivable, for
instance, that the equality of inertial and gravitational mass
break down for antimatter, or in quantum field theory at finite
temperatures (Donoghue {\it et al.}, 1984, 1985). Einstein's
equivalence principle is also violated in the Quantum Geometry
model developed by Caianiello as a first step toward the
unification of Quantum Mechanics and General Relativity. The model
interprets quantization as curvature of the eight-dimensional
space-time tangent bundle $TM_8$. In this space the standard
operators of the Heisenberg algebra are represented as covariant
derivatives and the quantum commutation relations are interpreted
as components of the curvature tensor.

In this paper we have analyzed the oscillation phenomena of
neutrinos propagating in a Schwarzschild geometry modified by the
existence of MA, which implies a violation of the equivalence
principle. We have calculated the quantum mechanical phase showing
that, for the consistence of the Caianiello model, our results are
compatible with estimations of the neutrino masses  giving
$m_\nu\sim 0.05\div 1$eV.

Eqs. (\ref{phasemot1}) and (\ref{phaseMA1}) allow to calculate the
flavor oscillation probability, which is given by
\begin{equation}\label{probability}
  P_{\nu_e\to \nu_\mu}=\sin^22\theta\sin^2\left(\pi\frac{\Delta
  r}{\lambda_{{\cal A}_m}}\right)\,,
\end{equation}
where $\theta$ is the mixing angle, and $\lambda_{{\cal A}_m}$ is
the oscillation length defined as (for simplicity we use the
natural units $\hbar =c=1$)
\begin{equation}\label{osc}
  \lambda_{{\cal A}_m}^{-1}=\frac{\Delta m^2}{4E\pi}+
  \frac{E^3}{24\pi}\frac{\Delta
  m^2(m_1^2+m_2^2)}{m_1^4m_2^4}\frac{r_s^2(r_A^2+r_Ar_B+r_B^2)}{r_A^3r_B^3}\,.
\end{equation}
As well known, in the cases of interest, the oscillation length
$\lambda$ does depend on the energy of neutrinos as
$\lambda^{-1}\sim E^n$ (Fogli {\it et al.}, 1999). Then
$\lambda_{{\cal A}_m}^{-1}$ corresponds to standard oscillation
plus the equivalence principle violation induced by the existence
of MA, $(n=-1)\oplus (n=3)$. The behaviour $\lambda^{-1}\sim
E^{-1}$ coming from a flavor depending coupling to gravitational
field, as proposed by Gasperini, Leung and Halprin (Gasperini,
1988, 1989; Halprin and Leung, 1991), appeared to fit the
SuperKamiokande data, as well as the other alternative mechanisms
(Barger {\it et al.}, 1999; Foot {\it et al.}, 1998; Chobey and
Goswami, 2000). Nevertheless a different analysis of such data,
including for example upward-going muons events, has been
performed in Refs. (Fogli {\it et al.}, 1999; Lipari and
Lusignolo, 1999). In these papers, it is shown that the best fit
does confirm, at least for atmospheric neutrinos, the standard
scenario as the dominant oscillation mechanism, whereas the
equivalence principle violation, as formulated in (Gasperini,
1988, 1989; Halprin and Leung, 1991), do not provide a viable
description of data.

Unlike the mechanism proposed in (Gasperini, 1988, 1989; Halprin
and Leung, 1991), in this paper we have suggested an alternative
mechanism for introducing, in the framework of Quantum Geometry, a
violation of the equivalence principle in the neutrino oscillation
physics. The main consequence of this approach, as shown in Eq.
(\ref{osc}), is a different behaviour of the inverse of the
oscillation length as function of the energy ($\sim E^3$) with
respect to that one obtained in (Gasperini, 1988, 1989; Halprin
and Leung, 1991) whose energy dependence has the functional form
$(E\Delta f \phi)^{-1}$, where $\phi$ is the constant
gravitational field and $\Delta f$ the measure of the violation of
the equivalence principle.

%Even though many efforts have been done till now for solving the
%neutrino oscillation problem, a definitive solution is far to be
%achieved. Much more studies are necessary for understanding the
%origin of neutrino masses, the mixing of states, and the analysis
%of collaborations involving in neutrino oscillations experiments.
%Only the future generation of experiments could provide new data
%for probing the $E^3$--dependence induced by MA corrections.
%allowing to establish whether the violation of the equivalence
%principle discussed in this paper occurs, and as a consequence,
%whether the Quantum Geometry model proposed by Caianiello is a
%concrete step towards an unified theory of Quantum Mechanics and
%General Relativity.

\bigskip

\centerline{\bf Acknowledgements}

Research supported by MURTS fund PRIN 99.

\bigskip

\leftline{\bf REFERENCES}

\noindent Ahluwalia, D.V., and Burgad, C., 1996, {\it Gen. Rel.
           Grav.} {\bf  28}, 1161;  gr-qc/9606031. \\
Barger, V., Learned, J.G., Pakvasa, S., and Weiler, T.J.,
                1999, {\it Phys. Rev. Lett.} {\bf 82}, 2640. \\
Bhattacharya, T., Habib, S., and Mottola, E., 1999, {\it Phys.
                  Rev.} {\bf D59}, 067301. \\
Bilenky, S.M. and Pontecorvo, B., 1978, {\it Phys. Rep.} {\bf
                  41}, 225. \\
Caianiello, E.R., 1981, {\it Lett. Nuovo Cimento} {\bf 32}, 65. \\
Caianiello, E.R., De Filippo, S., Marmo, G., and Vilasi, G.,
                  1982, {\it Lett. Nuovo Cimento} {\bf 34}, 112.
                  \\
Caianiello, E.R., Gasperini, M., and Scarpetta, G., 1990a,
                      {\it Il Nuovo Cimento} {\bf 105B}, 259. \\
Caianiello, E.R., Feoli, A., Gasperini, M., and Scarpetta,  G,
                1990b, {\it Int. J. Theor. Phys.} {\bf 29}, 131. \\
Caianiello, E.R., Gasperini, M, and Scarpetta, G., 1991,
                {\it Class. Quant. Grav.} {\bf 8} 659. \\
Caianiello, E.R., 1992, {\it Rivista del Nuovo Cimento} {\bf 15},
                  No. 4. \\
Capozziello, S., Lambiase, G., and Scarpetta, G.,
             1999, {\it Nuovo Cimento} {\bf 114B}, 93. \\
Capozziello, S., Feoli, A., Lambiase, G., Papini, G., and
                  G. Scarpetta, (2000a)
               {\it Phys. Lett.} {\bf A268}, 247.\\
Capozziello, S., Lambiase, G., Scarpetta, G., (2000b), {\it Int.
                 J. Theor. Phys.} {\bf 39}, 15. \\
Chen, C.X., Papini, G., Mobed, N., Lambiase, G., and
               Scarpetta, G., 1999, {\it Il Nuovo Cimento} {\bf B114},
               199.\\
Chen, C.X., Papini, G., Mobed, N., Lambiase, G., and
               Scarpetta, G., 2000, {\it Il Nuovo Cimento} {\bf B114}
               1335.\\
Choubey, S., and Goswami, S., 2000, {\it Astropart. Phys.} {\bf
                  14}, 67. \\
Donoghue, J.F., Holstein, B.R., and Robinett, R.W., 1984, {\it
                  Phys. Rev.} {\bf D30}, 2561. \\
Donoghue, J.F., Holstein, B.R., and Robinett, R.W., 1985, {\it
                  Gen. Rel. Grav.} {\bf 17}, 207.
Feoli, A., 1993, {\it Nucl. Phys.} {\bf B396}, 261. \\
 Feoli, A.,
Lambiase, G., Papini, G., and Scarpetta, G., 1997. {\it
               Il Nuovo Cimento} {\bf 112B} 913. \\
Feoli, A., Lambiase, G., Papini, G. and Scarpetta, G.,
               1999a, {\it Phys. Lett.} {\bf A263}, 147. \\
Feoli, A., Lambiase, G.,  Nesterenko, V.V., and
                Scarpetta, G. 1999b, {\it Phys. Rev.} {\bf D60},
                065001.\\
Fogli, G.L., Lisi, E., Marrone, A., and Scioscia, G., 1999,
               {\it Phys. Rev.} {\bf D60}, 053006. \\
Foot, R., Leung, C.N., and Yasuda, O., 1998, {\it Phys. Lett.}
                 {\bf B443}, 185. \\
Frolov, V.P., Sanchez, N., 1991, {\it Nucl. Phys.} {\bf B349},
                    815. \\
Gasperini, M., 1988, {\it Phys. Rev.} {\bf D38}, 2635. \\
Gasperini, M. 1989, {\it Phys. Rev.} {\bf D39}, 3606. \\
Gasperini, M., Sanchez, N., and Veneziano, G., 1991, {\it Nucl.
               Phys.} {\it B364} 365. \\
Gasperini, M., 1992, {\it Gen. Rel. Grav.} {\bf 24} 219. \\
Gibbons, G.W. and Hawking, S.W., 1977, {\it Phys. Rev.} {\bf D15},
                  2738. \\
Halprin, A. and Leung, C.N., 1991, {\it Phys. Rev. Lett.} {\bf
                  67}, 1833. \\
Kuwata, S., 1996, {\it Il Nuovo Cimento} {\bf B111}, 893. \\
Lambiase, G., Papini, G., and Scarpetta, G., 1998, {\it Phys.
               Lett.} {\bf A244}, 349. \\
Lambiase, G., Papini, G. and Scarpetta, G., 1999, {\it Il Nuovo
                Cimento} {\bf 114B}, 189. \\
Lipari, P. and Lusignoli, M., 1999, {\it Phys. Rev.} {\bf D60},
              013003. \\
McGuigan, M., 1994, {\it Phys. Rev.} {\bf D50}, 5225. \\
Mikhyev,
S.P., and Smirnov, A.Yu, 1986a, {\it Sov. J. Nucl. Phys.}
              {\bf 42}, 913. \\
Mikhyev, S.P., and Smirnov, A.Yu, 1986b, {\it Nuovo Cim.} {\bf
              C9}, 17. \\
Misner, C.W., Thorne, K.S., and Wheeler, 1973, J.A.,{\it
             Gravitation},
             W.H. Freeman and Co. San Francisco. \\
Papini, G., Feoli, A., and Scarpetta, G., 1995, {\it Phys. Lett.}
              {\bf A202} 50. \\
Sanchez, N., and Veneziano, G., 1990, {\it Nucl. Phys.} {\bf B333}
                253. \\
Sanchez, N., 1993, {\it ``Structure: from Physics to General
                Systems''}, Eds. M. Marinaro and G. Scarpetta,
                World Scientific, Singapore, vol. 1, pag. 118.\\
Stodolski, L, 1979, {\it Gen. Rel. Grav.} {\bf 11}, 391. \\
Wolfenstein, L., 1978, {\it Phys. Rev.} {\bf D17}, 2369.

%\end{thebibliography}

\newpage

\begin{center}
{\bf Table I:} Quantum mechanical phase mixing for atmospheric
neutrinos with fixed value of $\Delta m^2$ and $E\sim 1$GeV. $m_1$
and $m_2$ are expressed in eV.
\end{center}
\begin{center}
\begin{tabular}{ccccc}\hline\hline
  % after \\: \hline or \cline{col1-col2} \cline{col3-col4} ...
  $m_1$ & $m_2$ & $\Delta m^2$ & $\Delta\Phi_{(0)}$ & $\Delta\Phi_{{\cal A}_m}$
  \\ \hline\hline
  0.5 & 0.51 & $10^{-2}$ & $2.5\cdot 10^{-1}$ & $10^{-8}$ \\
  0.1 & 0.14 & $10^{-2}$ & $2.5\cdot 10^{-1}$ & $8.6\cdot 10^{-5}$ \\
  0.05 & 0.11 & $10^{-2}$ & $2.5\cdot 10^{-1}$ & $1.8\cdot 10^{-3}$ \\
  0.01 & 0.1 & $10^{-2}$ & $2.5\cdot 10^{-1}$ & 1.14 \\ \hline
  0.5 & 0.501 & $10^{-3}$ & $2.5\cdot 10^{-2}$ & $ 10^{-9}$  \\
  0.1 & 0.104 & $10^{-3}$ & $2.5\cdot 10^{-2}$ & $2\cdot 10^{-5}$  \\
  0.05 & 0.06 &  $10^{-3}$ & $2.5\cdot 10^{-2}$ & $8.9\cdot
  10^{-4}$ \\
  0.01 & 0.03 &  $10^{-3}$ & $2.5\cdot 10^{-2}$ & 1.13\\ \hline\hline
\end{tabular}
\end{center}

\vspace{0.5cm}

\begin{center}
{\bf Table II:} Quantum mechanical phase mixing for solar
neutrinos. Here $m_1\sim m_2\sim m$ are expressed in eV and $E$ in
MeV.
\end{center}
\begin{center}
\begin{tabular}{cccc}\hline\hline
  % after \\: \hline or \cline{col1-col2} \cline{col3-col4} ...
  $m$ & $E$ & $\Delta\Phi_{(0)}/\Delta m^2$ & $\Delta\Phi_{{\cal A}_m}/\Delta m^2$
  \\ \hline\hline
  0.5 & 1 & $2.5\cdot 10^{11}$ & $10^{2}$ \\
  0.1 & 1 & $2.5\cdot 10^{11}$ & $2\cdot 10^{6}$ \\
  0.05 & 1 & $2.5\cdot 10^{11}$ & $1.2\cdot 10^{8}$ \\
  0.01 & 1 & $2.5\cdot 10^{11}$ & $2\cdot 10^{12}$ \\ \hline
  0.5 & 10 & $2.5\cdot 10^{10}$ & $ 10^{5}$  \\
  0.1 & 10 & $2.5\cdot 10^{10}$ & $2\cdot 10^{9}$  \\
  0.05 & 10 & $2.5\cdot 10^{10}$ & $1.2\cdot 10^{11}$ \\
  0.01 & 10 & $2.5\cdot 10^{10}$ & $2\cdot 10^{15}$ \\ \hline\hline
\end{tabular}
\end{center}

\end{document}